# Levelized Cost of Energy for PV and Grid Scale Energy Storage Systems

Chun Sing Lai and M.D. McCulloch

Energy and Power Group, Department of Engineering Science, University of Oxford, UK

Corresponding Author Email: chun.lai@eng.ox.ac.uk

*Abstract*-- **With the increasing penetration of renewable energy sources and energy storage devices in the power system, it is important to evaluate the cost of the system by using Levelized Cost of Energy (LCOE). In this paper a new metric, Levelized Cost of Delivery (LCOD) is proposed to calculate the LCOE for the energy storage. The recent definitions in LCOE for renewable energy system has been reviewed. From fundamental principles, it is demonstrated that there is a need to introduce a new method to evaluate LCOE of the system as the conventional LCOE is not applicable for renewable energy storage systems. Three years of solar irradiance data in Africa collected from Johannesburg and the national load data from Kenya are obtained for case studies. The proposed cost calculation methods are evaluated for two types of storage technologies (Vanadium Redox Battery (VRB) and Lithium-ion) with real-life data. It shows that the marginal LCOE and LCOD indices can be used to assist policymakers to consider the discount rate and the type of storage technology for a cost effective renewable storage energy system.**

*Index Terms*-- **PV, Micro-grid, LCOE, LCOD, Energy Storage**

## I. INTRODUCTION

AS renewables take a larger share of generation capacity and where electrical systems cannot keep up with the increasing demand, increasing system flexibility should thus become a priority for policy and decision makers. Energy storage systems (ESS) could provide services and improvements to power grid systems, so storage may one day be ubiquitous in the power systems [1]. It is believed that energy storage will be a key asset in the evolving smart grid.

Successful operation of electric grid requires continuous real-time balancing of supply and demand including losses. As ESS options become increasingly available and countries around the globe continue to enrich their portfolios of renewable energy, the use of energy storage is increasing. For example, increased deployment of ESS in the distribution grid could make this process more effective and could improve system performance. Mainly, ESS mediates between variable sources and variable loads; works by moving energy through time. Essentially, ESS can smooth out this variability and allow electricity to be dispatched at a later time. ESSs are highly adaptable and can meet the needs of various users including renewable energy generators, grid equipment, and end users [2].

Industrial and digital economy firms are collectively losing $45.7 billion a year due to outages. These data suggest that across all business sectors, the US economy is losing between $104 billion and $164 billion a year due to outages and another $15 billion to $24 billion due to power quality phenomena [3]. By using ESS, the security of supply and power quality issue could potentially be minimized, with a reduction in outages.

There are many ways to calculate the economic viability of distributed generation and energy efficiency projects. The capital cost of equipment, the operation and maintenance costs, and the fuel costs must be combined in some ways so that a comparison may be made. One of the most commonly used metrics is the levelized cost of energy (LCOE).

In this paper, a PV renewable energy system with storage is used to demonstrate the marginal LCOE and LCOD methodologies due to the constraints in solar energy. Variable renewable generators (primarily wind, solar photovoltaics, and concentrating solar power when deploying without storage) are unlike conventional generators, they cannot be dispatched (except by curtailing output) and their output varies depending on local weather conditions, which are not completely predictable. In practice, the LCOE calculation method in this paper could be applied to any power generation sources, such as diesel generator.

Existing papers have given reasons for deployment of ESS in the future power system [4-6]. Many literatures analyzed the life cycle or levelized cost solely for storage component, without considering the cost at a system level and energy exchange between generation source and storage [7-10]. LCOE analysis for renewable systems (such as PV and wind energy) is also already well established and presented in many literatures [11-12]. However, cost analysis for such integrated system has not been given a proper treatment and have not been clearly justified.

A detailed review of recent LCOE calculation methods for renewable and ESS systems has been given and possible shortcomings of existing methods have been highlighted. A proposed LCOE (system) and LCOD has been derived from first principles. Practical solar irradiance, load, and the most recent system components cost data from literatures have been used for the analysis in this paper. The results have been compared with different sources to support the credibility of the proposed method. It is known that levelized cost of storage (LCOS) and LCOD will be higher than LCOE.

The methods present in this paper have filled in the knowledge gap in the cost calculations aspect for power systems with renewable energy sources and ESS. The paper proceeds as follows. The definition of LCOE will be reviewed in Section II. Section III will provide a survey in the recent trend of large-scale PV systems and the LCOE for renewable systems with storage devices. Section IV provides the derivation for the LCOD for the ESS and the LCOE (System) renewable system with ESS. Section V provides the case studies for calculations of marginal LCOE and LCOD. A real-life case study with the daily national load data of Kenya and 3 years of collected solar irradiance data from Johannesburg is given. Discussions and conclusions are given in Section VI and Section VII respectively.

## II. Levelized Cost of Energy

LCOE is a measure of costs which attempts to compare different methods of electricity generation on a comparable basis. It is an economic assessment of the average total cost to build and operate a power-generating asset over its lifetime divided by the total energy output of the asset over that lifetime. The LCOE can also be regarded as the minimum cost at which electricity must be sold in order to achieve break-even over the lifetime of the project. The aim of LCOE is to give comparison of different technologies (e.g., wind, solar, natural gas) of unequal life spans, project size, different capital cost, risk, return, and capacities.

The general equation for LCOE [13-14] is given in Equation (1). It is essentially the lifecycle cost of the system be divided by the lifetime energy production of the system.

$$LCOE = \frac{\text{Lifecycle cost (\$)}}{\text{Lifetime energy production (kWh)}} \quad (1)$$

As reported by Allan *et al.* [15], there are two methods commonly used to calculate the levelized costs, known as the "discounting" method, and the "annuitizing" method. In the discounting method shown in Equation (2), the stream of (real) future costs and (electrical) outputs identified as $C_t$ and $E_t$ in year t are discounted back with discount rate r, to a present value (PrV). The PrV of costs is then divided by the PrV of lifetime output. The levelized costs measured under the "discounting" method, $LCOE_{Discount}$, is given by

$$LCOE_{Discount} = \frac{\text{PrV(Costs)}}{\text{PrV(Output)}} = \frac{\sum_{t=0}^{n} \frac{C_t}{(1+r)^t}}{\sum_{t=0}^{n} \frac{E_t}{(1+r)^t}} \quad (2)$$

In the "annuitizing" method as shown in Equation (3), the present value of the stream of costs over the device's lifetime (including pre-development, construction, operation, and any decommissioning costs) is calculated and then converted to an equivalent annual cost, using a standard annuity formula. This equivalent annual cost is then divided by the average annual electrical output over the lifetime of the plant, where n is the lifetime of the system in years.

$$LCOE_{Annuitizing} = \frac{\text{Ann(Costs)}}{\text{Ave(Output)}}$$
$$= \frac{\left(\sum_{t=0}^{n} \frac{C_t}{(1+r)^t}\right)\left(\frac{r}{1-(1+r)^{-n}}\right)}{(\sum_{t=1}^{n} E_t)/n} \quad (3)$$

The two methods give the same levelized costs when the discount rate used for discounting costs and energy output in Equation (2) is the same as that used in calculating the annuity factor in Equation (3). However, for levelized costs to be the same under both measures, annual energy output must also be constant over the lifetime of the device. The annuity method converts the costs to a constant flow over time. This is appropriate where the flow of energy output is constant. It is commonly assumed in the literature on levelized cost estimates that annual energy output is constant. However, the annual energy output of renewable technologies would typically vary from day-to-day mainly due to variations in the renewable resources. Therefore, it is more appropriate to use the discounting method than the annuitizing method when calculating LCOE for renewable sources.

As mentioned by Branker *et al.* [16], one of the misconceptions when calculating LCOE is that the summation does not start from t = 0 to include the project cost at the beginning of the first year. The first year of the cost should not be discounted to reflect the present value and there is no system energy output to be degraded.

It is common to store the excess energy generated by PV in storage systems to be used later on. The PV energy in an available time instance should be used directly to support the load and to avoid the losses due to round trip efficiency. In splitting the total energy produced by the PV system into two types, known as the surplus and the direct energy. Surplus energy, $E_{surplus}$, is the extra energy generated by PV system and not consumed by the load. Direct energy, $E_{direct}$, is the energy that consumed by the load directly. $C_{pvsurplus}$ and $C_{pvdirect}$ are the costs for generate surplus energy and direct energy respectively.

$$LCOE_{PV} = \frac{C_{pvsurplus} + C_{pvdirect}}{E_{pvsurplus} + E_{pvdirect}} \quad (4)$$
$$LCOE_{PV} = \frac{C_{pvsurplus}}{E_{pvsurplus} + E_{pvdirect}}$$
$$+ \frac{C_{pvdirect}}{E_{pvsurplus} + E_{pvdirect}} \quad (5)$$

In Equation (5), it is proven that both terms are not in-line with the definition of LCOE. Hence, the LCOE definition for a system with energy delivery sources such as energy storage requires a different definition.

Ashuri *et al.* [11] presents a method for multidisciplinary design optimization of offshore wind turbines at system level. The objective function to be minimized is the levelized cost of energy, however, energy storage was not considered in this paper. Mandelli et al. [17] has commented that LCOE has been employed as an objective function in a number of analyses that deal with renewable-based off-grid systems. They modified the traditional definition of the LCOE by considering the internalization of the value of lost load-related costs. Diaz *et al.* [12] introduced time of installment as major component in the computation of the LCOE. Whereas the classic LCOE is static, i.e the installment is done today, the proposed methodology dynamically searches a point in the future where LCOE would be optimum. The papers have made a contribution to re-modify the usage of LCOE, it is worth noting that the storage has not been considered in the system.

Branker *et al.* [16] has also provided a review on the methodology of properly calculating the LCOE for solar PV. The equation for calculating the LCOE for a PV system is given in Equation (6) below:

$$
\begin{aligned}
LCOE &= \frac{\sum_{t=0}^{n}(I_t + O_t + M_t + F_t)/(1+r)^t}{\sum_{t=0}^{n}E_t/(1+r)^t} \\
&= \frac{\sum_{t=0}^{n}(I_t + O_t + M_t + F_t)/(1+r)^t}{\sum_{t=0}^{n}S_t(1-d)^t/(1+r)^t}
\end{aligned}
\tag{6}
$$

It should be noted that the initial investment $I_t$ is an one-off payment. It should not be discounted and it should be taken out of the summation. The LCOE for PV systems given by the author also considers the degradation factor of PV modules. The energy generated in a given year $E_t$ is the rated energy output per year $S_t$ multiplied by the degradation factor $(1-d)$ which decreases the energy with time. The maintenance costs, operation costs and interest expenditures for time year t are denoted as $M_t$, $O_t$ and $F_t$ respectively.

## III. LARGE-SCALE PV SYSTEM AND ENERGY STORAGE

### A. The System

There are a number of reasons why large-scale PV system will be the future direction and in order to promote this, many researchers have considered different scenarios to achieve this. Peters *et al.* [18] provide a comparative assessment of the three leading large-scale solar technologies in 2010 and 2020 for different locations. Also mentioned by Reichelstein *et al.* [19], it concludes that today these technologies cannot yet compete with conventional forms of power generation but approach competitiveness around 2020 in favorable locations. In addition, none of the solar technologies emerges as a clear winner and that costs of storing energy differ from technologies and can change the order of competitiveness in some instances. It should be noted that the competitiveness of the different technologies varies considerably across locations due to differences e.g., solar resource and discount rates. In order to further refine policy recommendations, some areas for future research are especially promising. Policymakers are in the need for more precise advice on which policy mixes are most workable to improve the usage of different technologies. Future research should assist policymakers in exploiting this potential by evaluating in more detail the needs for accompanying measures in the areas of storage and grid management.

Wang *et al.* [20] compared the LCOE across PV systems with equal installation areas but with modules of different efficiencies installed with fixed tilt, 1-axis tracking or 2-axis tracking. The first finding was that at a given module price in $/W, more efficient PV modules lead to lower LCOE systems. The second finding was that when meeting an LCOE goal, the PV module efficiency has a lower limit that cannot be offset by module price; and the third and final finding was that both 1-axis and 2-axis tracking installations provide lower LCOEs than fixed tilt installations.

Schill *et al.* [4] provide an investigation on renewable surplus generation and storage requirements in Germany. Surplus energies are generally low, but there are high surplus power peaks. It states that there are several questions remain for future research, in particular regarding the optimal mix of storage, curtailment and other flexibility options. The study of different energy storage technologies interaction with network expansion, power-to-heat, and thermal plants appears to be a particularly promising field of research. Additionally, the full system value of storage technologies should be investigated, including their capacity value and the provisions of ancillary services.

A simulation environment has been developed by Wade *et al.* [5] that enables multiple types of network event to be monitored and acted on simultaneously. Network conditions are recreated from historical data by load flow analysis and an assessment is made on the applicability of an intervention from ESS. The simulations conducted within the simulation environment have shown that operating an ESS embedded in the distribution network has a positive impact on the tasks of voltage control and power flow management. A higher power rating and energy capacity ESS could solve a greater number of problems but there is a balance of cost/benefit to be achieved. Electrical energy storage is one of the tools that will become increasingly available to network planners and operators. As progress is made in the transition to future electricity networks, electrical energy storage embedded at distribution level is set to become an integral part of the Smart Grid.

Denholm *et al.* [6] examined the changes to the electric power system required to incorporate high penetration of variable wind and solar electricity generation in a transmission constrained grid. The main issues on incorporation of these sources at large-scale are the limited time coincidence of the resource with normal electricity demand, combined with the limited flexibility of thermal generators to reduce output. These constraints would result in unusable renewable generation and increased costs. But a highly flexible system – with must-run base-load generators virtually eliminated – allows for penetrations of up to about 50% variable generation with curtailment rates of less than 10%. For penetration levels up to 80% of the system's electricity demand, keeping curtailments to less than 10% requires a combination of load shifting and storage equal to about one day of average demand.

It is impractical to install an ESS that is capable of providing a solution to all events at all times; either the events would have to be very modest or the ESS is very large. The ESS operates to make a contribution to improve network performance in cooperation with other Smart Grid control actions such as active generator curtailment or demand side management. The proportion of contribution made by energy storage depends upon the event schedule and the varying behavior of the network on both short and long time-scales.

Garvey *et al.* [21] has suggested energy storage and generation must be separated. There is an increasing acceptance that energy storage will play a major role in future electricity systems to provide at least a partial replacement for the flexibility naturally present in fossil-fuelled generating stations. It mentioned that if UK are to be powered completely by onshore wind turbines with large energy stores capable of delivering the exact required total energy, 30.3% of all energy consumed would have passed through storage. It had also mentioned that if all UK power come from PV with storage, 57.1% of all energy consumed would have passed through storage. Evidently, if future electricity systems are powered largely from inflexible sources, substantial fractions of all electrical energy consumed may pass through storage.

Gomez *et al.* [22] has given an overview of the Spanish power generation sector. The sector is facing dire problems: generation overcapacity, various tariff hikes over recent years, uncertainty over the financial viability of many power plants and a regulatory framework that lacks stability. They find that appropriate energy planning could have reduced investments in the Spanish power sector by 28.6 billion euro by 2010 without compromising on performance in terms of sustainability or energy security, while improving affordability. The main causes of these surplus investments were two supply factors: those of gas combined cycles and of solar technologies. The results of this work highlight the value of rigorous, quantitative energy planning, and the high costs of not doing it. ESS system could potentially improve the situation by reducing the required generation capacity by providing flexibility to the system.

Orioli *et al.* [23] has provided an economic analysis of the investment in grid-connected PV systems installed on the rooftops of buildings located in densely urbanized contexts. The LCOE (levelized cost of electricity) was calculated as an indicator of the competitiveness of the PV technology. The paper concludes that by considering the effects of the analyzed promoting schemes on the discounted payback period and LCOE, also considering the role of the solar energy shading and mismatch between electricity generation and consumption. The results presented in this paper seem to demonstrate that the new promoting policy may yield greater economic advantages for the domestic investors who have planned to install PV systems in urbanized areas. Unfortunately, although the competitiveness of the PV LCOE with retail electricity prices is an appealing goal, the trajectory towards the grid parity is still slow in Italy. The deployment of energy storage system could potentially improve the system's LCOE by providing energy balance.

### B. Storage and the Issues in Cost Calculation

Turning to energy storages, Poonpun *et al.* [7] has given a cost analysis of grid-connected electric energy storage. Various energy storage technologies are considered in the analysis. It calculates the cost of energy added by storing electricity for different storage technologies. It has made some comparisons solely for storage technologies and renewable energy system has not been considered in the paper.

Zakeri *et al.* [8] has highlighted that the economic implications of grid-scale electrical energy storage technologies are however obscure for the experts, power grid operators, regulators, and power producers. The paper has commented that if the cost of charging electricity would be deducted from the LCOE delivered by EES, the net levelized cost of storage (LCOS) is presented in Equation (7) [8].

$$LCOS = LCOE - \frac{\text{price of charging power}}{\text{overall efficiency}} \quad (7)$$

Equation (7) states that LCOS will be less than LCOE. The cost of storage should be higher than the cost of the system, storage cost needs to include the cost of energy generation to be stored in ESS. The storage will have an efficiency factor; hence the storage output energy will be lower than the energy generates by source. It is noted that the generation source in the calculation of LCOS or LCOE for the system has not been considered. The energy stored in storage system is affected by the energy production of renewable source.

In this paper, two types of energy storage systems will be studied, known as the Vanadium Redox Battery (VRB) and Lithium-ion battery (Li-ion).

In general, the future perspective seems to be promising for Li-ion batteries in grid-scale applications as the final price is declining and the functionality is ever improving by optimizing manufacturing costs, extending the lifetime, using new materials, and improving the safety parameters. Leadbetter *et al.* [9] identified that Lithium-ion and lead-acid batteries are suitable for short duration services. Whereas Sodium-sulfur and vanadium redox batteries are suitable for long duration services.

Singh *et al.* [10] has provided a study of the LCOE for hydrogen-bromine flow battery. It mentioned that although the capital cost of storage is an important and frequently reported method of evaluating battery cost, the most important metric is the levelized cost of electricity and the value that should be minimized, rather than minimizing capital cost. Currently, natural gas peaker plants evaluated with a lifetime of 20 years, and a capital cost of 1 $/Wpeak, at a natural gas cost of 4 $/MMBtu operating for 4 h per day would have a levelized cost of electricity of 0.14 $/kWh, not including any electronic infrastructure. At 0.40 $/kWh, the hydrogen-bromine flow battery system is too high for grid-level or any price-sensitive application. It is explained that the high cost of hydrogen storage is a major limitation of the hydrogen-bromine system compared to other flow battery systems, especially as it scales with energy storage capacity, unlike the stack which scales with power. The costs of the hydrogen-bromine system can be significantly lowered if the costs of the battery stack and power electronics can be reduced. Currently, the costs are competitive with other flow or stationary battery cell system, and thus can compete in the same markets. The performance of the battery, power density at a given efficiency also affects the levelized cost significantly, indicating that further improvements in the efficiency of the battery can have large influences on the cost of electricity. However, the large effect of a decrease in lifetime on the levelized cost of electricity indicates that durability of the system may be more important than minor improvements in performance.

According to an US Solar Energy Monitor report, lithium-ion batteries are the most common storage technology, regardless of application. Vanadium Redox batteries are emerging as another storage option. Lux Research reports that falling costs will lead to a 360-MWh market in 2020, worth $190 million. VRB is the most mature technology in the area of solar storage [24].

The World Energy Council [25] has proposed a new formula in Equation (8), known as the Levelized Cost of Storage to enable comparisons between different types of storage technologies in terms of average cost per produced / stored kWh.

$$LCOS = \frac{I_o + \sum_{t=1}^{n} \frac{C_{ESS_t}}{(1+r)^t}}{\sum_{t=1}^{n} \frac{E_{ESS_t}}{(1+r)^t}} \quad (8)$$

$I_o$ is the initial investment cost. $C_{ESS_t}$ and $E_{ESS_t}$ are the total costs and energy output at year t respectively. It is mentioned that the LCOS formula only summarizes the general LCOS of each technology, i.e. without applying the application cases for wind or PV systems. It shows that the renewables industry faces two main challenges when it applies the LCOS metric:

1. Arbitrariness: Storage levelized cost estimations are arbitrary, since the application case can vary widely.
2. Incompleteness: Storage levelized cost estimations are incomplete, since they do not cover the required business models and its characteristics for storage. In the LCOE philosophy, the required revenue is only reflected by the applied discount factor. Since it neglects higher potential revenues, e.g., from providing flexibility, it is a simplified approach for the actual value of storage.

The report has also emphasized that the energy sector has reasons to be enthusiastic about storage, but from the wrong perspective. Although investment cost reduction is important, there is a growing value of specific storage technologies in specific contexts. Policymakers should examine storage through holistic case studies in context, rather than only emphasis in generic cost estimations.

Lazard [26] modeled 10 different use cases for storage including frequency regulation, grid balancing and micro-grid support with the possibility of eight different storage technologies, ranging from compressed-air energy storage to lithium-ion batteries. The required energy output for different storage applications are predetermined. Because of the operating and physical conditions, some ESSs would need to be overrated. This oversizing results in depth of discharge over a single cycle that is less than 100%. While energy storage is a beneficiary of and sensitive to various tax subsidies, this report presents the LCOS on an unsubsidized basis to isolate and compare the technological and operational components of energy storage systems and use cases, as well as to present results that are applicable to a global energy storage market.

The LCOS provided by Lazard is an optimistic estimation and in practice, the storage system will not be used to 100% of its capacity. In the case of PV integration, the energy stored in the storage system depends on the PV system output and this is highly arbitrary as it depends on the nature of solar irradiance. Therefore, the LCOS will be different in real-life situation and expected to be higher. The values provided by Lazard can be used as a comparison for different storage technologies and applications, but cannot be used for system resource planning and decision making. The operational parameters used in Lazard's LCOS study for PV integration are presented in Table 1 below.

Table 1: Parameters used for LCOS study for PV integration [26]

| Parameters | |
|---|---|
| Project lifetime (Year) | 20 |
| Discount rate (%) | 8 |
| Storage power capacity (MW) | 2 |
| Storage energy capacity (MWh) | 4 |
| Cycles per day (100% DOD) | 1.25 |
| Days of operation per year | 350 |
| Annual energy production (MWh) | 1750 |
| System's total generate energy (GWh) | 35 |

The report did not provide the method for the LCOS calculation. It is assumed that the results are calculated with Equation (8). The results of LCOS for PV integration with the lower and upper bound range for different storage technology is provided in Table 2 below.

Table 2: Current LCOS for PV integration [26]

| | LCOS ($/kWh) | |
|---|---|---|
| Storage Type | Lower bound | Upper bound |
| Zinc | 0.245 | 0.345 |
| Vanadium battery (VRB) | 0.373 | 0.950 |
| Lithium-ion | 0.355 | 0.686 |
| Lead | 0.402 | 1.068 |
| Sodium | 0.379 | 0.957 |

Pawel [27] has provided a new methodology for the calculation of levelized cost of stored energy. Many new terms have been proposed such as price increase factor and internal transfer cost to calculate the LCOE of the hybrid system. In practice, it is difficult to determine these values. Hence it is not a practical method to calculate the LCOE.

Mundada et al. [28] reported that economic projections on complex hybrid systems utilizing a combination of PV, battery and cogeneration is challenging and no comprehensive method is available for guiding decision makers. These authors claimed to have provided a new method of quantifying the economic viability of off-grid PV/battery/CHP systems by calculating the levelized cost of electricity (LCOE) of the technology to be compared to centralized grid electricity. The most important factors for determining LCOE of hybrid system are system cost, financing, operation and maintenance cost, fuel cost, loan-term and lifetime. The authors did not specify the importance of the energy production and capacity factor. It is important to know that the energy output from PV depends on the degradation rate of the modules.

Proposed LCOE for the hybrid system is given in Equation (9) [28] below:

$$LCOE = \frac{I + \sum_{t=1}^{n} \frac{(I * i + O + F_{chp})}{(1+r)^t}}{\sum_{t=1}^{n} E_{tpv}(1-d_1)^t + \frac{E_{tchp}(1-d_2)^t}{(1+r)^t}} \quad (9)$$

I is the total installation cost which includes the cost of solar PV, battery and the CHP module, i is the interest rate on the hybrid system for 100% debt financing. O is the total operation and maintenance cost. $F_{chp}$ is the annual fuel cost of the CHP unit. $E_{tpv}$ and $E_{tchp}$ are the rated annual energy production from solar PV and CHP unit respectively. $d_1$ and $d_2$ are the degradation rates for solar PV and CHP unit respectively. The energy produced by PV system is not discounted. It does not reflect the actual value of the solar PV energy in the future. Cost implication due to storage has not been included in the analysis in detail. Although storage does not generate energy, the total energy production will be affected by storage due to round trip efficiency. The total energy production by the system is therefore inaccurate in this study.

## IV. LCOD DERIVATION

This section of the paper provides the explanation on the reason why the conventional LCOE definition is an issue for renewable and storage systems.

Figure 1 shows the energy flow diagram of the renewable energy and storage system. The total PV panels in the system are divided into two sets. One set is to generate the energy for the ESS while the other set is to supply the load directly. The net energy output of the ESS needs to take account of the round trip efficiency η.

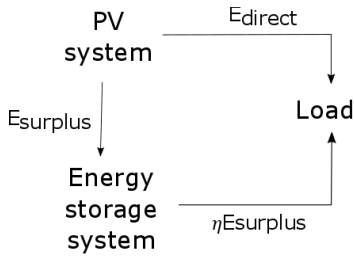

Figure 1: Energy flow diagram of the system

### A. Cost calculations for storage system

Figure 2 shows the direction of energy flow of ESS. The derivation of the LCOE for the ESS is given in Equations (10-13). The LCOE of the energy into the system is given in Equation (10).

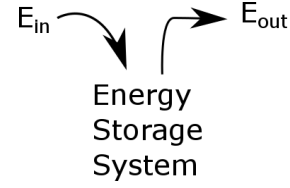

Figure 2: Energy flow in and out of the ESS

$$LCOE(E_{in}) = \frac{\sum_{t=0}^{n} \frac{C_{in_t}}{(1+r)^t}}{\sum_{t=0}^{n} \frac{E_{in_t}}{(1+r)^t}} \quad (10)$$

$C_{in_t}$ is the total cost for delivering the PV energy into ESS at year t. $E_{in_t}$ is the input energy to the ESS at year t. The LCOE for ESS in a renewable energy system is more complicated to comprehend. It is necessary to take the cost of the solar panels to generate the surplus energy to be stored into the ESS into account. This is due to the fact the energy stored in the ESS is produced by the solar panels. The LCOE of the energy delivered by the ESS is given in Equations (11) and (12) below:

$$LCOE(E_{out}) = \frac{\sum_{t=0}^{n} \frac{C_{in_t}}{(1+r)^t} + \sum_{t=0}^{n} \frac{C_{ESS_t}}{(1+r)^t}}{\eta \sum_{t=0}^{n} \frac{E_{in_t}}{(1+r)^t}} \quad (11)$$

$$LCOE(E_{out}) = \frac{\sum_{t=0}^{n} \frac{C_{in_t}}{(1+r)^t}}{\eta \sum_{t=0}^{n} \frac{E_{in_t}}{(1+r)^t}} + \frac{\sum_{t=0}^{n} \frac{C_{ESS_t}}{(1+r)^t}}{\eta \sum_{t=0}^{n} \frac{E_{in_t}}{(1+r)^t}} \quad (12)$$

By splitting Equation (11) into two individual components, the final form of the LCOE for the ESS is given in Equation (13).

$$LCOE(E_{out}) = LCOD = \frac{1}{\eta} LCOE(E_{surplus}) + LCOS \quad (13)$$

In practice, $E_{in}$ will be the surplus energy, $E_{surplus}$ flowing into the storage to be a dispatchable source of power. Therefore, $C_{in}$ will be $C_{PVsurplus}$, the PV panels that produced the surplus energy for the system.

### B. Systems with Renewable and Storage

For a PV and ESS power system, the following LCOE relationship will hold:

$$LCOE_{system} = \frac{\sum_{t=0}^{n} \frac{C_{system_t}}{(1+r)^t}}{\sum_{t=0}^{n} \frac{E_{system_t}}{(1+r)^t}} \quad (14)$$

$C_{system_t}$ and $E_{system_t}$ are the total cost and total energy

production from the system at time t respectively. The total cost of the renewable system is the sum of PV generation and storage costs. The total energy produced by the system is the energy output of ESS and the energy directly delivered to the load by PV. Therefore, the LCOE for the system is given in Equation (15) below.

$$LCOE_{system} = \frac{C_{pvsurplus} + C_{ESS} + C_{pvdirect}}{E_{ESS} + E_{pvdirect}} \qquad (15)$$

As induced from Equations (4) and (5), it should be noted that the definition of LCOE should be modified for this case. Since the energy "delivered" in this case is not the energy "generated" by the system.

## V. CASE STUDIES

### A. Simulations

#### 1) Cost and Energy Calculation

This section presents the cost calculation of the system and analysis of the results. Table 3 gives the cost specification of the components for the energy system.

The solar panel to be used for the system is the Sharp ND-R250A5. It has an efficiency of 15.3% and has a rated power of 250W [29]. Systems with Vanadium Redox battery and Lithium-ion battery are studied respectively. The ESS power rating is 2MW with a capacity of 4MWh.

Table 3: Cost and technical specification of the system components

| | PV (Sharp ND-250QCS) | ESS | |
| | | Vanadium redox battery (VRB) | Lithium-ion battery |
|---|---|---|---|
| Capital cost ($C_{Cap}$) | 120 ($/unit) [29-30] | 760-1600 ($/kWh) [29,31] | 715-1640 ($/kWh) [29,31] |
| Installation cost ($C_{Inst}$) | 108 ($/unit) [29] | N/A | N/A |
| O&M cost ($C_{O&M}$) | 6 ($/unit/year) [29] | 100-140 ($/kWh) [29] | 80-95 ($/kWh) [29] |
| System Lifetime (n) | N/A | 20 years[29,31] | 15 years [29,31] |
| Round trip efficiency (η) | N/A | ~70% [29,31] | ~90% [9,29,31] |

The total cost and the energy production from the PV system and ESS for the LCOE calculations are given in Equations (16) to (20) below:

$$C_{ESS} = C_{Cap\_ESS} + \sum_{t=0}^{n} \frac{C_{O\&M\_ESS_t}}{(1+r)^t} \qquad (16)$$

$$E_{ESS} = \eta \sum_{t=0}^{n} \frac{E_{surplus_t}(1 - D_{ESS})^t}{(1+r)^t} \qquad (17)$$

$$C_{PVsurplus} = (C_{Cap\_pv} + C_{Inst\_pv} + \sum_{t=0}^{n} \frac{C_{O\&M\_pv_t}}{(1+r)^t}) N_{PVsurplus} \qquad (18)$$

$$C_{PVdirect} = (C_{Cap\_pv} + C_{Inst\_pv} + \sum_{t=0}^{n} \frac{C_{O\&M\_pv_t}}{(1+r)^t}) N_{PVdirect} \qquad (19)$$

$$E_{PVdirect} = \sum_{t=0}^{n} \frac{(E_{direct} * 365)_t (1 - D_{pv})^t}{(1+r)^t} \qquad (20)$$

In this study, the storage system degradation rate, $D_{ESS}$ is at 1% [27] while PV panel degradation rate, $D_{pv}$ is at 0.5% [32-33]. $N_{pvsurplus}$ and $N_{pvdirect}$ are the number of units of PV panels for producing surplus and direct energy respectively.

As reported in [34], the current discount rate for Solar PV is 6-9%. The discount rate could be as much as 2-3% lower over the next decade, and could fall by a further 1-2% by 2040.

#### 2) Marginal Levelized Cost of Energy Calculation

Marginal cost plays a key role in the economic theory that proves a competitive market is efficient, but there are also two practical uses of marginal cost that increase its importance in a power market. Firstly, many power markets rely on a central day-ahead auction in which generators submit individual supply curves and the system operator uses these to determine the market price. Because price should equal marginal cost in an efficient market, the auction rules should be informed by a coherent theory of marginal cost. Secondly, many power markets suffer from potential market-power problems which cause the market price to diverge from marginal cost [35].

The definition of marginal cost is the cost of producing one more unit of output. In practice, load power and solar power are highly non-linear due to the energy consumption habit of consumers and the diurnal effect with weather perturbation respectively. To keep the problem simple for the study, it is assumed that the load curve is constant and fixed for all cases. The solar curve produced from the solar irradiance model [36] is at clear-sky condition. In this paper, three different cases of the marginal cost of LCOE are studied with the following assumptions.

Case 1: The peak of solar power meets the load demand.

Case 2: The solar capacity increases by 1.5 times, the surplus energy will be discarded because of no storage.

Case 3: Storage will be used to store the surplus energy. Number of the PV panel is the same as that in Case 2.

In this paper, three different cases of marginal costs are studied and it should be noted that $C_{case(k)}$ is the annual cost for the PV panels and $E_{pv(k)}$ is the daily energy production from the PV panels in case k. The total cost and energy to calculate the LCOE for the system in the case studies are given in Equations (21) to (23) below:

$$C_{total\_case(k)} = \sum_{t=0}^{n} \frac{C_{case(k)_t}}{(1+r)^t} \quad (21)$$

$$E_{total\_case(k)} = \sum_{t=0}^{n} \frac{(E_{pv(k)} * 365)_t}{(1+r)^t} \quad (22)$$

The energy $E_{pv(k)}$ is the daily energy production and it needs to be multiplied by 365 days to provide the annual energy production. In Case 1, the load uses all the produced solar energy. LCOE is then calculated. Figure 3 shows the visual representation of Case 1.

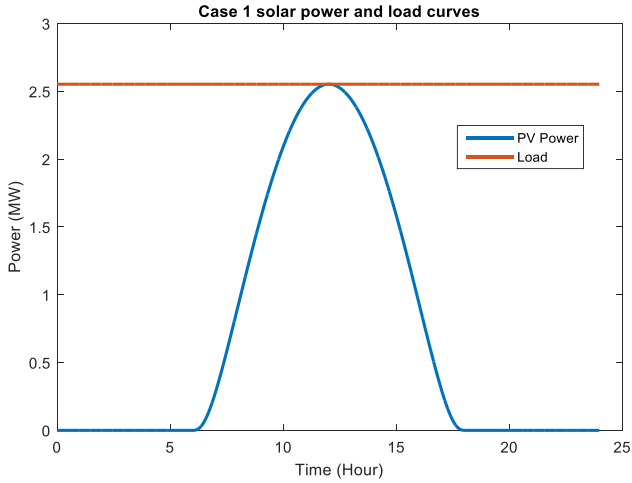

Figure 3: Solar and load curve for Case 1

The load power is assumed to be at the maximum point of the solar power curve for the default case. There are 20000 panels at 15.3% efficiency. The LCOE for the default case is:

$$LCOE_{basecase} = \frac{C_{total\_case1}}{E_{total\_case1}} \quad (23)$$

In Case 2, assuming that 10000 extra panels are invested into the system or that the number of the panels have been increased by 1.5 times from Case 1. However, there is no storage device. Therefore, the surplus energy will be wasted. The shaded area is the extra solar energy produced in the system that consumed by the load compared to Case 1. Figure 4 shows the visual representation of Case 2. $E_{PV2}$ is the annual energy production from total PV panels in Case 2. $E_{total\_case2}$ is obtained by taking away $E_{surplus}$ from the total energy production from PV, $E_{PV2}$ due to no energy storage present in the system.

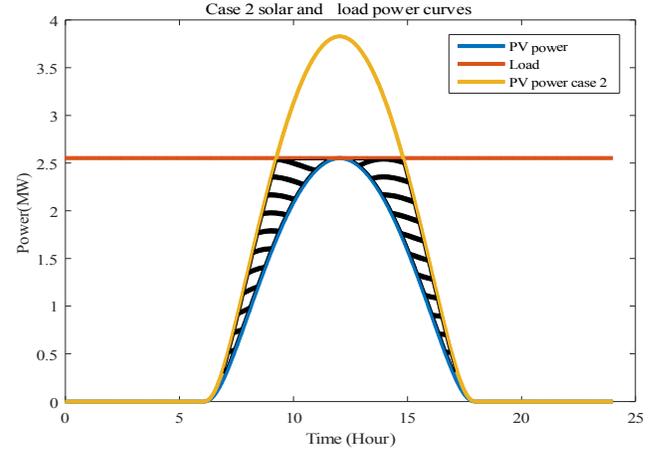

Figure 4: Solar and load curve for Case 2

The marginal LCOE from Case 1 to Case 2 is:

$$LCOE_{marginal(1-2)} = \frac{\Delta C}{\Delta E} = \frac{C_{total\_case2} - C_{total\_case1}}{E_{total\_case2} - E_{total\_case1}} \quad (24)$$

$$\text{where } E_{total\_case2} = (E_{PV2} - E_{surplus}) \quad (25)$$

In Case 3, further investment is put into the system as compared to Case 2 by including ESS. The surplus energy will be stored in the ESS and consumed by the load. The surplus energy Esurplus is 4.676 MWh per day. The rated capacity of ESS is 5MWh. Figure 5 shows the visual representation of Case 3. $E_{pv3}$ is the annual energy production from all solar panels in the Case 3. The marginal LCOE from Case 2 to Case 3 is:

$$LCOE_{marginal(2-3)} = \frac{\Delta C}{\Delta E} = \frac{C_{total\_case3} - C_{total\_case2}}{E_{total\_case3} - E_{total\_case2}}$$

$$= \frac{(C_{ESS} + C_{total\_case2}) - C_{total\_case2}}{E_{total\_case3} - E_{total\_case2}}$$

$$= \frac{C_{ESS}}{E_{ESS}} \quad (26)$$

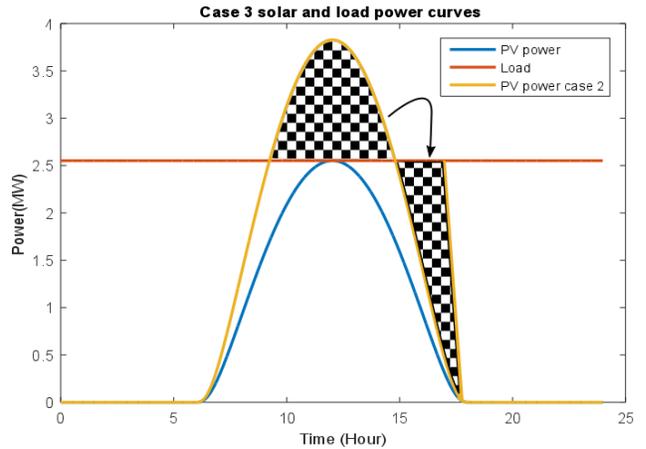

Figure 5: Solar and load curve for Case 3

The marginal LCOE from Case 1 to Case 3

$$LCOE_{marginal(1-3)} = \frac{\Delta C}{\Delta E}$$
$$= \frac{(C_{total\_case2} - C_{total\_case1}) + C_{ESS}}{(E_{total\_case2} - E_{total\_case1}) + E_{ESS}} \quad (27)$$

Table 4: LCOE and marginal LCOE cases for VRB at lower bound cost

| r (%) | LCOE$_{basecase}$ ($/kWh) | LCOE$_{marginal}$ ($/kWh) | | | LCOD ($/kWh) | LCOE$_{system}$ ($/kWh) |
| | | 1-2 | 2-3 | 1-3 | | |
|---|---|---|---|---|---|---|
| 2 | 0.095 | 0.207 | 0.168 | 0.186 | 0.400 | 0.154 |
| 5 | 0.108 | 0.236 | 0.212 | 0.223 | 0.460 | 0.173 |
| 8 | 0.121 | 0.265 | 0.259 | 0.262 | 0.525 | 0.193 |
| 10 | 0.130 | 0.286 | 0.291 | 0.289 | 0.570 | 0.207 |
| 15 | 0.154 | 0.337 | 0.374 | 0.355 | 0.685 | 0.242 |

Table 5: LCOE and marginal LCOE cases for Lithium-ion at lower bound cost

| r (%) | LCOE$_{basecase}$ ($/kWh) | LCOE$_{marginal}$($/kWh) | | | LCOD ($/kWh) | LCOE$_{system}$ ($/kWh) |
| | | 1-2 | 2-3 | 1-3 | | |
|---|---|---|---|---|---|---|
| 2 | 0.102 | 0.224 | 0.153 | 0.183 | 0.344 | 0.156 |
| 5 | 0.115 | 0.251 | 0.182 | 0.212 | 0.386 | 0.173 |
| 8 | 0.127 | 0.279 | 0.214 | 0.242 | 0.430 | 0.191 |
| 10 | 0.136 | 0.298 | 0.235 | 0.263 | 0.461 | 0.204 |
| 15 | 0.158 | 0.345 | 0.291 | 0.315 | 0.539 | 0.235 |

The LCOD given in Table 4 is in range with the Lazard's result of LCOS in Table 2. The following observations could be made.

1. System without storage attracts a smaller LCOE but naturally at a higher risk of security of supply. (column 1)
2. From marginal cost, it can be seen that energy waste will lead to a higher LCOE, so it would be important to add a battery to minimize energy wastage and to potentially reduce the LCOE.
3. From the new method, it can be seen that it is important to add a battery as a component of the system rather than adding it in a later stage. The earlier addition can lead to a smaller LCOE.

### B. Practical case study with Africa data

The purpose of this study is to calculate the LCOE for a practical renewable energy system with storage. A scenario has been developed with the load and generation data from Africa. Two types of dominant storage technologies lithium-ion battery VRB are analysed. Three years of completed solar irradiance data in 2009, 2011 and 2012 were collected in Johannesburg for the practical case study. The sampling rate is at 30 min/sample. Figure 7 shows the annual solar power generation from a 5MW PV farm for 2009, 2011 and 2012.

The national load curve of Kenya is presented in Figure 6. As explained in [37], the national peak starts building at 18:30 and attains its peak at 20:30. The load peak then starts reducing gradually at 20.30 when the domestic load is shut down. At 22:00, the domestic load is completely shut down leaving only the few industrial and commercial consumers run for 24 hours. The shape of the national load profile for South Africa and Kenya are very similar [38-39]. It is therefore assumed that the load curve for Johannesburg is similar to that for Kenya. The peak of the load curve is around 8pm and the lowest consumption is around 5am in the morning. For the case study in this paper, the national load curve has been down-sized with the peak load at 2MW.

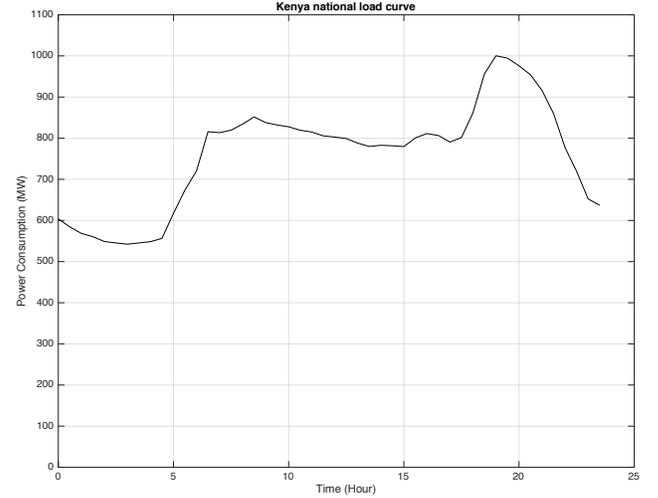

Figure 6: Daily national load curve in Kenya

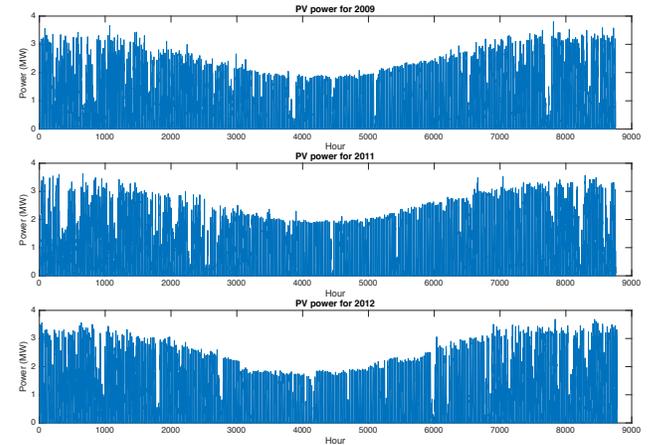

Figure 7: Annual solar power curve with solar farm rated at 5MW

The LCOD and the LCOE$_{system}$ were given in Equations (13) and (15) respectively. As explained in this paper, the cost of the energy delivered by storage needs to take into account of the solar panel for producing the surplus energy. Equations (28) and (29) determine the amount of PV panels used for producing the potential maximum amount of energy for storage and direct consumption respectively. The storage capacity is 5MWh. $\sigma$ is at 15% panel efficiency, $\varepsilon$ is the solar irradiance at W/m2, m is the number of hours per year.

$$N_{direct} = \frac{\int_{t=0}^{m} P_{direct}(t)}{\sigma \int_{t=0}^{m} \varepsilon(t)} \quad (28)$$

$$N_{surplus} = \frac{\int_{t=0}^{m} P_{surplus}(t)}{\sigma \int_{t=0}^{m} \varepsilon(t)} \qquad (29)$$

The results of LCOD and LCOE$_{system}$ are given below:

Table 6: Case study with Lithium-ion at lower bound cost

| d (%) | LCOD ($/kWh) | | | LCOE$_{system}$ ($/kWh) | | |
|---|---|---|---|---|---|---|
| | 2009 | 2011 | 2012 | 2009 | 2011 | 2012 |
| -2 | 0.417 | 0.401 | 0.371 | 0.175 | 0.171 | 0.166 |
| 0 | 0.457 | 0.439 | 0.407 | 0.189 | 0.185 | 0.180 |
| 2 | 0.499 | 0.480 | 0.444 | 0.204 | 0.200 | 0.194 |
| 5 | 0.567 | 0.545 | 0.505 | 0.228 | 0.224 | 0.217 |
| 8 | 0.638 | 0.614 | 0.569 | 0.253 | 0.248 | 0.241 |
| 10 | 0.687 | 0.661 | 0.613 | 0.270 | 0.265 | 0.257 |
| 15 | 0.813 | 0.782 | 0.726 | 0.313 | 0.307 | 0.298 |

Table 7: Case study with VRB at lower bound cost

| d (%) | LCOD ($/kWh) | | | LCOE$_{system}$ ($/kWh) | | |
|---|---|---|---|---|---|---|
| | 2009 | 2011 | 2012 | 2009 | 2011 | 2012 |
| -2 | 0.462 | 0.443 | 0.410 | 0.169 | 0.165 | 0.161 |
| 0 | 0.516 | 0.495 | 0.459 | 0.184 | 0.181 | 0.176 |
| 2 | 0.575 | 0.553 | 0.512 | 0.201 | 0.197 | 0.192 |
| 5 | 0.673 | 0.646 | 0.599 | 0.228 | 0.223 | 0.217 |
| 8 | 0.778 | 0.748 | 0.693 | 0.256 | 0.251 | 0.244 |
| 10 | 0.850 | 0.818 | 0.758 | 0.275 | 0.270 | 0.263 |
| 15 | 1.037 | 0.997 | 0.925 | 0.324 | 0.318 | 0.310 |

Table 8: Case study with Lithium-ion battery at upper bound cost

| d (%) | LCOD ($/kWh) | | | LCOE$_{system}$ ($/kWh) | | |
|---|---|---|---|---|---|---|
| | 2009 | 2011 | 2012 | 2009 | 2011 | 2012 |
| -2 | 0.684 | 0.656 | 0.605 | 0.231 | 0.226 | 0.219 |
| 0 | 0.764 | 0.733 | 0.677 | 0.253 | 0.248 | 0.240 |
| 2 | 0.849 | 0.815 | 0.753 | 0.276 | 0.271 | 0.263 |
| 5 | 0.986 | 0.946 | 0.874 | 0.312 | 0.306 | 0.297 |
| 8 | 1.130 | 1.084 | 1.001 | 0.350 | 0.343 | 0.333 |
| 10 | 1.229 | 1.180 | 1.089 | 0.376 | 0.369 | 0.358 |
| 15 | 1.482 | 1.423 | 1.314 | 0.441 | 0.432 | 0.420 |

Table 9: Case study with VRB at upper bound cost

| d (%) | LCOD ($/kWh) | | | LCOE$_{system}$ ($/kWh) | | |
|---|---|---|---|---|---|---|
| | 2009 | 2011 | 2012 | 2009 | 2011 | 2012 |
| -2 | 0.690 | 0.662 | 0.611 | 0.211 | 0.207 | 0.201 |
| 0 | 0.792 | 0.760 | 0.701 | 0.234 | 0.230 | 0.223 |
| 2 | 0.904 | 0.868 | 0.801 | 0.259 | 0.254 | 0.247 |
| 5 | 1.088 | 1.044 | 0.964 | 0.299 | 0.293 | 0.285 |
| 8 | 1.285 | 1.234 | 1.139 | 0.341 | 0.334 | 0.325 |
| 10 | 1.422 | 1.365 | 1.261 | 0.369 | 0.362 | 0.353 |
| 15 | 1.772 | 1.701 | 1.571 | 0.441 | 0.433 | 0.421 |

The LCOD and LCOE$_{system}$ are dissimilar every year due to different annual energy production by the PV system. The amount of PV power is directly proportional to the solar irradiance received. It can be observed that the LCOD is cheaper for Lithium-ion battery in both extremes.

From Tables 6 and 7 for lower bound cost case at 2% discount rate, LCOE$_{system}$ is cheaper for VRB. There is a crossover point when the discount rate is at 5%. Lithium-ion is cheaper for discount rate above 5%.

The reason for LCOE$_{system}$ VRB is less than LCOE$_{system}$ Lithium-ion when discount rate is below 5% because energy decreases dramatically with increased discount rate. This will also increase the LCOE for both storage technologies.

From Tables 8 and 9 for upper bound cost case, the LCOE$_{system}$ for VRB is cheaper than lithium-ion in all cases. The LCOE$_{system}$ approaches crossover point at 15%. It is expected with the further increased discount rate; lithium-ion will be cheaper. The crossover increases to 15% is due to the capital cost being more dominant over the effect of operation and maintenance cost.

In the future, it is very likely that the discount rate will be less than 5% so VRB will be a potential good choice from the economical point of view.

## VI. Discussions

A more accurate calculation of LCOS has been given in this paper, known as the LCOD. The method has been confirmed with practical data and critical analysis has been given. With the increasing amount of storage in the energy systems, it is crucial to analyze the economic values to determine the feasibility of such systems. This method could also be used to assess different storage technologies although in the paper, only VRB and lithium-ion battery cases were given as examples. This method provides decision makers with a practical approach to consider the competitiveness of each technology for a given application with renewables in particular. From the analysis, it can be seen that LCOD and LCOE$_{system}$ are different for each year. Main parameters such as efficiency, lifetime, discount rate have all been included.

Although the LCOD is cheaper for Lithium-ion than that for VRB at present, the system LCOE could be lower for VRB when the discount rate is less than 5%.

Since energy storage has many applications for power systems such as grid balancing and frequency regulation, the LCOS and LCOD will be significantly different due to the operating conditions of the ESS. The future work would be to analyze the storage costs for different applications. With more irradiance data, it will be possible to have a better determination of the cost in deployment of renewables system with integration of energy storage system. In the future, state of charge and discharge and cycles could be considered.

## VII. Conclusions

This paper has provided a review of LCOE and highlighted the recent developments in large-scale solar PV systems. From the basic principles, the LCOD and the LCOE of renewable energy systems with storage has been proposed. The results conclude that in the future VRB could potentially substitute Lithium-ion for the energy storage application in PV systems due to the lower LCOE and the falling discount rates for renewable energy systems.

## VIII. Acknowledgement

The authors would like to thank Prof Ken Nixon, University of Witwatersrand for providing the irradiance data for this research. Load data from Kenya provided by Dr Mike Mason, Tropical Power is also appreciated.